\PassOptionsToPackage{unicode,hypertexnames=false}{hyperref}
\documentclass{ws-mpla}

\usepackage[english]{babel}

\usepackage[utf8]{inputenc}
\usepackage[super]{cite}

\newcommand{\eql}[1]{\label{eq:#1}}
\newcommand{\eq}[1]{(\ref{eq:#1})}

\usepackage{amsmath, amssymb}
\usepackage{mathrsfs}
\DeclareMathAlphabet{\mathsl}{OT1}{cmr}{m}{sl}

\newcommand{\ud}{\textrm d}
\newcommand{\scalprod}[2]{\left\langle #1,#2 \right\rangle}
\newcommand{\La}{\mathscr{L}}
\newcommand{\M}{\mathcal{M}}
\newcommand{\Or}[1][1]{\mathcal O(#1)}

\newcommand{\g}[2][]{\mathsl{g}^{#1}_{#2}}
\newcommand{\e}[2][]{e^{#1}_{#2}}
\newcommand{\he}[2][]{\hat e^{#1}_{#2}}
\newcommand{\gb}[2][]{\g[#1]{\pmb{#2}}}
\newcommand{\eb}[2][]{\e[#1]{\pmb{#2}}}
\newcommand{\heb}[2][]{\he[#1]{\pmb{#2}}}

\newcommand{\fv}[2][]{\varphi^{\,#1}_{#2}}
\newcommand{\fs}[2][]{\widetilde\varphi^{\,#1}_{#2}}
\newcommand{\fvb}[2][]{\fv[#1]{\pmb{#2}}}
\newcommand{\fsb}[2][]{\fs[#1]{\pmb{#2}}}

\AtBeginDocument{
  \hypersetup{pdftitle={Vector boson scattering and boundary conditions in KK toy model},pdfauthor={P. Morávek \& J. Hořejší}}
}

\begin{document}

\markboth{P. Morávek \& J. Hořejší}
{Vector boson scattering and boundary conditions in KK toy model}

\catchline{}{}{}{}{}

\title{VECTOR BOSON SCATTERING AND BOUNDARY CONDITIONS IN KALUZA-KLEIN TOY MODEL}

\author{PETR MORÁVEK\footnote{Corresponding author} , JIŘÍ HOŘEJŠÍ}

\address{Institute of Particle and Nuclear Physics,\\
Faculty of Mathematics and Physics, Charles University,\\
V Holesovickach 2, 180 00 Prague 8, Czech Republic\\
\footnotemark[1] moravek@ipnp.troja.mff.cuni.cz}

\maketitle

\pub{Received (22 February 2012)}{}

\begin{abstract}
We study a~simple higher-dimensional toy model of electroweak symmetry breaking, in~particular a~pure gauge 5D theory on flat background with one extra finite space dimension.
The~principle of least action and the~requirement of gauge independence of~scattering amplitudes are used to determine the~possible choices of boundary conditions.
We demonstrate that for any of these choices the~scattering amplitudes of vector bosons do not exhibit power-like growth in the~high energy limit.
Our analysis is an~extension and generalization of the~results obtained previously by other authors.

\end{abstract}

\ccode{PACS Nos.: 11.10.Kk, 11.15.Bt, 14.70.-e, 14.80.Rt}

\section{Introduction}
Electroweak symmetry breaking (EWSB), i.e. the~mechanism of generating the~$W$ and $Z$ boson masses, is one of the~most important theoretical issues of the~present-day particle physics.
Several viable scenarios are available in the~current literature (for a~review, see e.g. Ref.~\citen{ewsb-review}) and it is clear that only experiments can resolve this long-standing puzzle.
In this respect, we are in a~rather fortunate situation now, since the~first preliminary results from LHC experiments are already coming and we can expect some important hints to the nature of EWSB in the~horizon of one year.

A~simple way of implementing the~EWSB is the~``textbook'' Higgs mechanism that leads inevitably to one or several elementary scalar bosons in the~physical spectrum.
While the~obvious paradigm for such a~scheme is the~current standard model (SM), there are other highly popular theories built along these lines: most notably, models involving supersymmetry have been intensely studied during the~last two decades or so, since they alleviate the~famous hierarchy problem (i.e. that of stabilizing the~scale of the~Higgs boson mass) considered by many to be a~technical flaw of the~SM.\cite{hierarchy-problem}
Needless to say, models with elementary scalars are most convenient from the~calculational point of view, since they are perturbatively renormalizable.
It also means that the~tree-level scattering amplitudes are unitarized automatically in the~high energy limit if the~Higgs particles are not too heavy.\cite{CLT,horejsi-unitarity,LQT}

Taking into account the~hierarchy problem, a~radical alternative would be a~model with no Higgs scalars at all.
The~oldest example of such a~higgsless version of the~EWSB is the~technicolor and its various ramifications (cf. Ref.~\citen{technicolor} for a~review), for which the~original conceptual paradigm is the~chiral symmetry breaking in QCD.
While such a~scheme is obviously quite attractive a~priori, the~application of the~ideas of dynamical symmetry breaking in the~area of electroweak interactions runs into specific difficulties and thus remains problematic so far. 

With the~advent of modern applications of the~higher-dimensional theories of~the~Kaluza-Klein~(KK) type, new attempts to attack the~EWSB problem have been made during the~last decade and models with compact extra dimensions have thus become increasingly popular (for a~review, see e.g. Refs.~\citen{extra-dimensions,csaki-tasi}).
A~particularly attractive scenario is EWSB via a~non-trivial choice of boundary conditions.
Although the~underlying higher-dimensional theory is non-renormalizable, the~unitarity breakdown is postponed to the~cutoff scale of the~effective 4D theory, which is related to the~size of the~extra dimension.
``Bad high energy behavior'' of scattering amplitudes is prevented by the~exchange of KK excitations rather than through elementary scalar particles.
These models thus belong to the~class of higgsless theories.

It is worth noting that the higher-dimensional theory can be viewed as a~limiting case of deconstructed 4D ``moose'' theory.\cite{deconstruction}
Such an approach recieved a~considerable attention including the formulation of KK equivalence theorem.\cite{kk-unitarity,deconstruction-without-bc,deconstruction-sum-rules}
Although the~deconstruction formalism can successfully restore the~higher-dimensional theory, it is still quite instructive to study the formulation of the effective 4D theory from higher dimensions and the scattering of vector bosons without relying on~the~KK equivalence theorem.
This approach has been already intensively studied (see in~particular Refs.~\citen{csaki-tasi,csaki-interval,kk-gf,kk-transformations}), but in our opinion there are still some points that need clarification, because the~results presented in the~literature so far are not sufficiently general and complete, even at the~level of simplified toy models.

We consider a~pure gauge theory on flat background with one extra finite space dimension and shortly review its construction.
We examine which choices of boundary conditions are allowed by the~principle of least action and the~requirement of gauge independence of scattering amplitudes in a~simple $SU(2)$ toy model.
We demonstrate that all of the~allowed choices lead to the~theory with scattering amplitudes that do not exhibit power-like growth in the~high energy limit.
In this way, previous results of other authors are extended and generalized.

\section{Gauge theories on an~interval}
We start with the~5D Yang-Mills Lagrangian on flat background, where the~extra space dimension is restricted to a~finite interval, conventionally denoted as $(0, \pi R)$.
This may be written\footnote{We use capital Latin letters for 5D Lorentz indices running through 0, 1, 2, 3, 5. Small Greek letters as usual stand for 4D Lorentz indices.} as
\begin{equation}
  \eql{lagrangian-gauge}
  \La_\text{gauge} = - \frac 1 4 F^a_{MN}F^{a MN} = - \frac 1 4 F^a_{\mu\nu}F^{a \mu\nu} - \frac 1 2 F^a_{\mu 5}F^{a \mu 5} \,\text{.}
\end{equation}

The $F^a_{\mu 5}F^{a \mu 5}$ part contains a~quadratic term mixing fields $A^a_\mu$ and $A^a_5$, but we can eliminate it by adding a~suitable gauge fixing term to the~Lagrangian.
Since the~compactification procedure generally breaks 5D Lorentz invariance, we do not need to limit ourselves to 5D invariant gauge fixing terms\cite{kk-gf} and are free to choose
\begin{equation}
  \eql{gf}
  \La_\text{g.f.} = - \frac 1 {2\xi} \left( \partial_\mu A^{a \mu} - \xi \partial_5 A_5^a \right)^2 \,\text{.}
\end{equation}

Such a~term is still invariant under the~usual 4D Lorentz transformations and exactly cancels the~cross term.
Furthermore, after the~KK expansion, which we perform later in the~Section~\ref{sec:effective-lagrangian}, the~part independent of $A^a_5$ agrees with the~usual Lorenz-type gauge fixing term for each KK mode of $A^a_\mu$ and the~propagators of vector modes have a~form known from R$_\xi$ gauge of the~Standard Model.
The~unitary gauge is given by the~limit $\xi \rightarrow \infty$.
All massive scalar modes are unphysical; they are eliminated in the~unitary gauge, playing a~similar role as the~would-be Goldstone bosons in the~Standard Model.

\section{Principle of least action}
When using the~variational principle of least action, we have to keep the~boundary terms coming from the~integration by parts in the~direction of the~extra space dimension.
Remember that it has a~finite length, so there is no reason to assume a~priori that the~fields (or their variations) vanish at the~endpoints of the~interval.
One thus gets
\begin{align}
  \delta S_\text{gauge} ={}& \int \ud^4 x \int_0^{\pi R} \ud y \, \Big[ \left( \partial_M F^{a M \nu} - \g{5} f^{abc} F^{b M \nu} A^c_M \right) \delta A^a_\nu + {}\notag\\
  &{}+ \left( \partial_\mu F^{a\mu 5} - \g{5} f^{abc} F^{b\mu 5} A^c_\mu \right)  \delta A^a_5 \Big] - \int \ud^4 x \Big[ F^{a 5 \nu} \delta A^a_\nu \Big]_0^{\pi R} \,\text{,}\\
  \delta S_\text{g.f.} ={}& \int \ud^4 x \int_0^{\pi R} \ud y \, \bigg[\left( \frac 1 \xi \partial^\nu\partial^\mu A^a_\mu - \partial^\nu\partial_5 A_5^a \right) \delta A^a_\nu + {}\notag\\
  &{}+ \left( \xi \partial_5\partial_5 A_5^a - \partial_5 \partial^\mu A^a_\mu \right) \delta A_5^a \bigg] + \int \ud^4 x \left[ \left( \partial^\mu A^a_\mu - \xi \partial_5 A_5^a \right) \delta A_5^a \right]_0^{\pi R} \,\text{.}
\end{align}
We thus have the~equations of motion plus some consistency conditions on the~fields at the~endpoints $0$ and $\pi R$.
Whatever boundary conditions we impose on the~gauge fields, we obviously need to ensure that the~two boundary terms vanish, i.e.
\begin{subequations}\eql{bc-restrict1}\begin{align}
  \left[ F^{a 5 \nu} \delta A^a_\nu \right]_0^{\pi R} ={}& 0 \,\text{,}\eql{bc1}\\
  \left[ \left( \partial^\mu A^a_\mu - \xi \partial_5 A_5^a \right) \delta A_5^a \right]_0^{\pi R} ={}& 0 \,\text{.}\eql{bc2}
\end{align}\end{subequations}

There are many possibilities how to satisfy \eq{bc-restrict1}.
The~least complicated way is to ensure that the~expressions vanish for every gauge field at each boundary separately, in other words require that the~variation itself or its coefficient is zero.
Assuming that we impose the~same boundary conditions for all colors of gauge fields (but we can impose different conditions at each endpoint of the~interval), we have three general choices of boundary conditions, namely
\begin{subequations}\eql{bc}\begin{align}
  && && A^a_\nu &= const & \text{and} && \partial_5 A^a_5 &= 0 \eql{bc-DN} \,\text{,} && &&\\
  && && \partial_5 A^a_\nu &= 0 & \text{and} && A^a_5 &= 0 \eql{bc-ND} \,\text{,} && &&\\
  && && A^a_\nu &= const & \text{and} && A^a_5 &= const \eql{bc-DD} \,\text{.} && &&
\end{align}\end{subequations}

The~last option simply means requiring that the~variations vanish ($\delta A^a_M = 0$) at the~boundary.
Later on we show that this choice leads to the~theory with $\xi$-dependent scattering amplitudes, thus we will omit this option from our discussion for the~moment.

Before proceeding with our examination of possible boundary conditions, we make several assumptions to simplify the~problem a~little:
\begin{itemlist}
  \item The~gauge group is $SU(2)$.
  \item Since we want to construct two charged bosons $W^\pm_M = \left( A^1_M \mp i A^2_M \right)/\sqrt{2}$ with the~same masses, we will always impose the~same boundary conditions on the~fields $A^1_M$ and $A^2_M$.
  \item We consider only the~Dirichlet ($\psi = 0$) and Neumann ($\partial_5 \psi = 0$) boundary conditions, thus we assume that all constants in \eq{bc} are zero.
\end{itemlist}

When reading Ref.~\citen{csaki-interval}, one could easily get an~impression that we can impose an~arbitrary combination of those boundary conditions at each endpoint of the~interval and for each color of gauge fields.
This is not quite true, because the~expression \eq{bc1} mixes fields of different colors, so when imposing different boundary conditions on different colors of gauge fields, we need to be sure that this term still vanishes.

The~above simplification leaves us with 16 different combinations of boundary conditions~-- two boundaries $\times$ two possible boundary conditions for each of two types of bosons, but seven (almost a~half) of them do not satisfy the~condition \eq{bc1}.
Those are the~cases with the~condition \eq{bc-ND} for $A^{1,2}_M$ and \eq{bc-DN} for $A^3_M$ at the~same endpoint of the~interval.
As an~explicite example let us write down the~expression \eq{bc1} for the~color $a = 1$:
\begin{equation}
  \left[ \partial^\nu A^1_5 - \partial_5 A^{1 \nu} - \g{5} \left( A^2_5 A^{3 \nu} - A^3_5 A^{2 \nu} \right) \right] \delta A^1_\nu = \g{5} A^3_5 A^{2 \nu} \delta A^1_\nu \neq 0
\end{equation}

\section{Effective 4D Lagrangian}\label{sec:effective-lagrangian}
To get the~effective 4D fields one can use the~KK expansion, i.e. decompose all fields into an~infinite series of eigenfunctions $\fv{a,n}(y)$ of the~operator $\partial_5\partial_5$.
The~decomposition of the~vector field $A^a_\mu(x,y)$ is then given by
\begin{equation}\eql{kk}
  A_\mu^a (x,y) = \sum_n A_\mu^{a,n}(x) \,\fv{a,n}(y)
\end{equation}
and the~$\fv{a,n}(y)$ is then called the~wave function (in the~extra dimension) of the~mode $A_\mu^{a,n}(x)$.
Similar decomposition can be done for scalar fields $A^a_5$ with a~different set of the~wave functions due to the~different boundary conditions.

As long as the~imposed boundary conditions keep the~operator $\partial_5\partial_5$ hermitian\footnote{One can easily check that the~Dirichlet or Neumann boundary conditions indeed keep the~operator hermitian.} with respect to the~scalar product $\scalprod f g = \int_0^{\pi R}\ud y\, f^{*}(y) g(y)$, we are guaranteed that its eigenfunctions satisfying
\begin{equation}\eql{eigenfunctions}
  \fv[\prime\prime]{a,n}(y) = -m_{a,n}^2\fv{a,n}(y)
\end{equation}
form a~complete orthonormal basis.
In this basis each mode $A_\nu^{a,n}$ of the~infinite KK tower obeys 4D equation of motion $\big(\square_4 + m_{a,n}^2 \big) A_\nu^{a,n}(x) - \partial_\nu \partial^\mu A^{a,n}_\mu(x) = 0$, thus it is effectively 4D vector boson with mass $m_{a,n}$.

We get the~effective 4D Lagrangian by means of a~simple integration over the~extra space dimension.
The~result\footnote{For the~sake of simplicity we use a~shorthand notation for multi-indices ${\pmb a} = (a, n)$.} can be split into several parts:
\begin{equation}
  S = \int \ud^4 x\, \left( \La_\text{free}^\text{vector} + \La_\text{free}^\text{scalar} + \La_\text{int}^\text{vector} + \La_\text{int}^\text{scalar} \right) \,\text{.}
\end{equation}
\begin{subequations}\eql{L4D1}\begin{align}
  \La_\text{free}^\text{vector} ={}& \frac 1 2 \sum_{\pmb a} A^{\pmb a}_\mu \,\square_4 A^{\pmb a \mu} + m_{\pmb a}^2 A^{\pmb a}_\mu A^{\pmb a \mu} + \Big(1 - \frac 1 \xi \Big) \Big(\partial_\mu A^{\pmb a \mu}\Big)^2 \eql{L4D-free-vector}\\
  \La_\text{free}^\text{scalar} ={}& \frac 1 2 \sum_{\pmb a} (\partial_\mu A^{\pmb a}_5)(\partial^\mu A^{\pmb a}_5) - \xi \widetilde m_{\pmb a}^2 A^{\pmb a}_5 A^{\pmb a}_5 \eql{L4D-free-scalar}\\
  \La_\text{int}^\text{vector} ={}& \sum_{\pmb{abc}} \gb{abc} f^{abc} A^{\pmb a}_\nu A^{\pmb b}_\mu \partial^\mu A^{\pmb c \nu} - \frac 1 4 \sum_{\pmb{abcd}} \gb[2]{abcd} f^{abe} f^{cde} A^{\pmb a}_\mu A^{\pmb c \mu} A^{\pmb b}_\nu A^{\pmb d \nu} \eql{L4D-int-vector}\\
  \La_\text{int}^\text{scalar} ={}& \sum_{\pmb{abc}} \heb{abc} f^{abc} A^{\pmb a}_5 A^{\pmb b}_\mu A^{\pmb c\mu} + \eb{abc} f^{abc} A^{\pmb a}_\mu A^{\pmb b}_5 \partial^\mu A^{\pmb c}_5 +{}\notag\\
    &{}+ \frac 1 2 \sum_{\pmb{abcd}} \eb[2]{abcd} f^{abe}f^{cde} A^{\pmb a}_\mu A^{\pmb c\mu} A^{\pmb b}_5 A^{\pmb d}_5 \eql{L4D-int-scalar}
\end{align}\end{subequations}
The~result contains the~free field Lagrangian for the~infinite tower of vector and scalar fields~-- each vector mode has the~Lorenz-type gauge fixing term and clearly all scalars are unphysical and are eliminated in the~unitary gauge (except if there is a~massless mode).
Further, there is an~interaction of vector bosons only, which has a~well known Yang-Mills structure and an~interaction that involves at least one scalar particle in every vertex.
All the~effective 4D couplings $\gb{abc}$, $\heb{abc}$, $\eb{abc}$ and $\gb[2]{abcd}$, $\eb[2]{abcd}$ are defined by an~integral of wave functions.
In our analysis we will need only three of them, explicitly
\begin{subequations}\eql{couplings}\begin{align}
  \gb[2]{abcd} &= \g[2]{5} \int_0^{\pi R} \ud y\, \fvb{a} \,\fvb{b} \,\fvb{c} \,\fvb{d} \eql{couplings-4V} \,\text{,}\\
  \gb{abc} &= \g{5} \int_0^{\pi R} \ud y\, \fvb{a} \,\fvb{b} \,\fvb{c} \eql{couplings-3V} \,\text{,}\\
  \heb{abc} &= \frac{\g{5}}{2} \int_0^{\pi R} \ud y\, \fsb{a} \,\left(\fvb{b} \,\fvb[\prime]{c} - \fvb[\prime]{b} \,\fvb{c} \right) \,\text{.} \eql{couplings-scalar}
\end{align}\end{subequations}

\section{Gauge independence of scattering amplitudes}
We require that the~scattering amplitude of process $VV \rightarrow VV$ does not depend on~the~gauge parameter $\xi$.
The~gauge parameter is present only in the~part of vector propagator that is proportional to $q_\mu q_\nu$ and the~mass of a~scalar field (thus also in its propagator), so the~only relevant (lowest order) diagrams are the~$s$, $t$ and $u$ channel exchange of a~scalar or vector particle.
The~gauge dependent terms must cancel out in each channel separately.
Let us take a~look at e.g. the~$s$ channel:
\begin{align}
  \M^{\text{(}s\text{)}}_\text{vector} = \sum_{\pmb e} &- \gb{eab} f^{eab} \left( m_{\pmb b}^2 - m_{\pmb a}^2 \right) \, \frac{1}{q^2 - m_{\pmb e}^2} \frac{1-\xi}{q^2 - \xi m_{\pmb e}^2} \, \gb{ecd} f^{ecd} \times {}\notag\\
  &\times \left( m_{\pmb d}^2 - m_{\pmb c}^2 \right) \left[\epsilon(k) \cdot \epsilon(l)\right] \left[\epsilon(p) \cdot \epsilon(r)\right] + \dots \,\text{,} \eql{R-vector}
\end{align}
\begin{equation}\eql{R-scalar}
  \M^{\text{(}s\text{)}}_\text{scalar} = \sum_{\pmb e} 2i \heb{eab} f^{eab} \, \frac{1}{q^2 - \xi \widetilde m_{\pmb e}^2} \, 2i \heb{ecd} f^{ecd} \left[\epsilon(k) \cdot \epsilon(l)\right] \left[\epsilon(p) \cdot \epsilon(r)\right] \,\text{.}
\end{equation}
In order to have any chance of cancellation between the~corresponding modes of exchanged vector and scalar particles, we obviously need $m_{\pmb e} = \widetilde m_{\pmb e}$.
This means that we need to impose either the~same boundary conditions on $A^a_5$ as on $A^a_\mu$, or the~opposite boundary conditions (meaning every Dirichlet condition imposed on a~vector field implies the~Neumann condition on the~scalar field of the~same color at the~same boundary and vice versa).

One can easily check that the~gauge dependent parts cancel out, if the~couplings and masses satisfy the~relation
\begin{equation}\eql{couplings-gg-ee}
  \gb{eab} \left( m_{\pmb b}^2 - m_{\pmb a}^2 \right) \gb{ecd} \left( m_{\pmb d}^2 - m_{\pmb c}^2 \right) = 2 \heb{eab} 2 \heb{ecd} m_{\pmb e}^2 \,\text{,}
\end{equation}
which can be recast in terms of the~integrals of wave functions as follows
\begin{multline}
  \g[2]{5} \int_0^{\pi R}\ud y\, \fvb[\prime]{e}\left(\fvb[\prime]{a}\,\fvb{b} - \fvb{a}\,\fvb[\prime]{b}\right) \int_0^{\pi R}\ud z\, \fvb[\prime]{e}\left(\fvb[\prime]{c}\,\fvb{d} - \fvb{c}\,\fvb[\prime]{d}\right) ={} \\
  {}= \g[2]{5} \,m_{\pmb e}^2 \int_0^{\pi R}\ud y\, \fsb{e}\left(\fvb[\prime]{a}\,\fvb{b} - \fvb{a}\,\fvb[\prime]{b}\right) \int_0^{\pi R}\ud z\, \fsb{e}\left(\fvb[\prime]{c}\,\fvb{d} - \fvb{c}\,\fvb[\prime]{d}\right) \,\text{.}\eql{couplings-gg}
\end{multline}

Let us examine the~relation between the~wave functions $\fvb{e}$ of the~vector modes and $\fsb{e}$ of the~scalar modes.
If we use the~boundary condition \eq{bc-DD}, then the~functions are the~same and obviously we cannot get gauge independent scattering amplitudes.
The~same conclusion was also reached in Ref.~\citen{kk-transformations} using a~different line of argumentation, based on the~requirement of consistently defined restricted class of 5D gauge transformations.
On the~other hand, if we use an~arbitrary combination of boundary conditions \eq{bc-DN} and \eq{bc-ND}, then one of the~functions is sine and the~other cosine with the~same arguments, and the~relation for every massive mode of color $e$ reads $m_{\pmb e}^2\,\fsb{e}(y)\,\fsb{e}(z) = \fvb[\prime]{e}(y)\,\fvb[\prime]{e}(z)$.

There could still be a~problem coming from massless modes, but since we have already established that the~only consistent boundary conditions are \eq{bc-DN} and \eq{bc-ND}, there can be only a~massless scalar, or a~massless vector particle, but not both of the~same color.
The~massless scalar is not a~problem, because the~gauge parameter is present only in the~term $\xi m_{e,0}^2 = 0$.
For a~massless vector boson, the~relevant term is proportional to the~expression \eq{couplings-gg}, which contains $\fv[\prime]{e,0} = 0$ under both integrals.
Since the~wave function of a~massless mode is a~simple constant, this term does not contribute to the~scattering amplitude at all.

The~conclusion of this section is that we can impose an~arbitrary combination\footnote{For completeness we should also show the~gauge independence of scattering amplitudes for processes with one or more massless scalars in the~initial or final state, but the~reasoning remains the~same, only the~integrals in coupling definitions are a~bit different.} of boundary conditions \eq{bc-DN} and \eq{bc-ND} on the~gauge fields as long as it satisfies \eq{bc-restrict1}.
This fact allows us to pass to the~unitary gauge, which simplifies significantly further calculations.

\section{Tree-level unitarity of $V_\text{L} V_\text{L} \rightarrow V_\text{L} V_\text{L}$ process}
Let us now calculate the~energy dependence of the~invariant matrix element for the~(generally inelastic) scattering of gauge bosons without any assumptions regarding the~color or the~KK mode number of the~gauge bosons in the~initial or final state.
We consider the~high energy limit and expand all quantities in powers of energy (more precisely in the~powers of Mandelstam invariant $s$) keeping only the~divergent parts, and show that they indeed cancel out automatically without introducing an~additional Higgs field.
We do not employ a~hard cutoff on the~spectrum of the~KK modes and keep the~whole infinite towers of the~KK excitations.
This is justified due to the~fact that the~contributions from the~highest KK modes are suppressed in the~high energy limit (for a~detailed discussion see e.g. Ref.~\citen{csaki-interval}).
The~lowest order diagrams for this process involve the~direct four-boson interaction, the~$s$, $t$ and $u$ channel exchange of KK vector excitations and possibly the~exchange of massless scalar in all channels as well.

We carry out the~calculation in the~center of mass reference frame.
Let us denote the~scattering angle by $\theta$ and for the~sake of simplicity introduce a~shorthand notation $4\overline m^2 = m_{\pmb a}^2 + m_{\pmb b}^2 + m_{\pmb c}^2 + m_{\pmb d}^2$.

The~energy expansion of the~contribution of the~contact four-boson interaction is given by
\begin{align}
  &\M^\text{(4V)} = \left(\frac s 4\right)^2 \frac{\gb[2]{abcd}}{m_{\pmb a} m_{\pmb b} m_{\pmb c} m_{\pmb d}} \bigg[ f^{eab}f^{ecd} (4\cos\theta) +{}\notag\\
  &+ f^{eac}f^{ebd} (-3 + 2\cos\theta + \cos^2\theta) + f^{ead} f^{ebc} (-3 - 2\cos\theta + \cos^2\theta)\bigg] +{}\notag\\
  &+ \left(\frac s 4\right)^{\phantom{2}} \frac{4\overline m^2 \gb[2]{abcd}}{m_{\pmb a} m_{\pmb b} m_{\pmb c} m_{\pmb d}} \bigg[ f^{eab}f^{ecd} (-\cos\theta) +{}\notag\\
  &+ f^{eac}f^{ebd} \frac{1 - \cos\theta}{2} + f^{ead}f^{ebc} \frac{1 + \cos\theta}{2} \bigg] + \Or \,\text{.} \eql{4V}
\end{align}

The~contributions of $s$, $t$, $u$ channel exchange of KK vector excitations are infinite sums over all KK modes\footnote{Note that due to the~structure of $SU(2)$ group the~color $e$ of exchanged gauge boson is fixed by the~colors of the~bosons in the~initial and final state, thus we actually sum only over the~KK index~$k$.}, which may be massive as well as massless.
Thus, with regard to the~different form of vector propagator for massive and massless modes, it is convenient to split them in two parts, namely $\M^{\text{(}s,t,u\text{)}}_\text{(long)}$ and $\M^{\text{(}s,t,u\text{)}}_\text{(diag)}$, corresponding to the~longitudinal ($q^\mu q^\nu$) and the~diagonal ($g^{\mu\nu}$)  parts of the~propagator respectively.
Since the~$t$ and $u$ channels differ only in the~simultaneous exchange of indices $\pmb c$ and $\pmb d$, and sign change of $\cos\theta$, from now on we will explicitly display only the~results for the~$s$ and $t$ channel.

Terms corresponding to the~longitudinal part of boson propagator can be expanded in the~powers of energy as follows:
\begin{subequations}\eql{st-qq}\begin{align}
  \M^{\text{(}s\text{)}}_\text{(long)} &= \sum_{k > 0} \frac s 4 \frac{\gb{eab}\gb{ecd} f^{eab}f^{ecd}}{m_{\pmb a} m_{\pmb b} m_{\pmb c} m_{\pmb d}} \frac{(m_{\pmb a}^2 - m_{\pmb b}^2)(m_{\pmb c}^2 - m_{\pmb d}^2)}{m_{\pmb e}^2} \, (-1) + \Or \,\text{,}\\
  \M^{\text{(}t\text{)}}_\text{(long)} &= \sum_{k > 0} \frac s 4 \frac{\gb{eac}\gb{ebd} f^{eac}f^{ebd}}{m_{\pmb a} m_{\pmb b} m_{\pmb c} m_{\pmb d}} \frac{(m_{\pmb a}^2 - m_{\pmb c}^2)(m_{\pmb b}^2 - m_{\pmb d}^2)}{m_{\pmb e}^2} \, \frac{1 - \cos\theta}{2} + \Or \,\text{.}
\end{align}\end{subequations}

Similarly, after quite a~long calculation, one gets the~terms corresponding to the~diagonal part of boson propagator in the~form
\begin{subequations}\eql{st-diag}\begin{align}
  \M^{\text{(}s\text{)}}_\text{(diag)} &= \sum_{k \geq 0} \frac{\gb{eab}\gb{ecd} f^{eab}f^{ecd}}{m_{\pmb a} m_{\pmb b} m_{\pmb c} m_{\pmb d}} \left[\left(\frac s 4\right)^2 (-4\cos\theta) + \frac s 4 (-m_{\pmb e}^2 \cos\theta)\right] + \Or \,\text{,}\\
  \M^{\text{(}t\text{)}}_\text{(diag)} &= \sum_{k \geq 0} \frac{\gb{eac}\gb{ebd} f^{eac}f^{ebd}}{m_{\pmb a} m_{\pmb b} m_{\pmb c} m_{\pmb d}} \bigg[ \left(\frac s 4 \right)^2 (3 - 2\cos\theta - \cos^2\theta) +{}\notag\\
  &\qquad\quad {}+ \left(\frac s 4\right) \Big(-m_{\pmb e}^2 \,\frac{3 + \cos\theta}{2} + 8\overline m^2 \cos\theta \Big)\bigg] + \Or \,\text{.}
\end{align}\end{subequations}

The~only scalar mode that can be present in the~theory is massless and the~corresponding contributions to the~invariant matrix elements read
\begin{subequations}\eql{st-scalar}\begin{align}
  \M^{\text{(}s\text{)}}_\text{(scalar)} &= \frac{s}{4}\, \frac{2\heb{eab}\, 2\heb{ecd} f^{eab}f^{ecd}}{m_{\pmb a} m_{\pmb b} m_{\pmb c} m_{\pmb d}} (-1) + \Or \,\text{,}\\
  \M^{\text{(}t\text{)}}_\text{(scalar)} &= \frac{s}{4}\, \frac{2\heb{eac}\, 2\heb{ebd} f^{eac}f^{ebd}}{m_{\pmb a} m_{\pmb b} m_{\pmb c} m_{\pmb d}} \frac{1 - \cos\theta}{2} + \Or \,\text{.}
\end{align}\end{subequations}
Owing to the~relation \eq{couplings-gg-ee} for all massive modes ($k > 0$) of color $e$ these terms give us in combination with $\M^{\text{(}s,t,u\text{)}}_\text{(long)}$ the~sum over the~complete orthonormal set of functions $\fs{e,k}$ in each channel.

It is a~matter of simple exercise to derive the~sum rule
\begin{equation}\eql{E4-condition}
  \gb[2]{abcd} = \sum_{k \geq 0} \gb{eab} \gb{ecd}
\end{equation}
which implies that terms growing as the~fourth power of energy in the~$s$, $t$ and $u$ channel contributions \eq{st-diag} cancel against the~terms from \eq{4V} corresponding to the~contact four-boson interaction.

Let us present some additional sum rules that are valid for all the~remaining consistent boundary conditions.
The~combination of contributions \eq{st-qq} and \eq{st-scalar} gives rise to the~sum of terms containing two couplings of the~type $\heb{eab}$.
The~index of KK mode is present only through these couplings, thus we can write down the~first sum rule
\begin{equation}\eql{E2-sum1}
  \sum_{k \geq 0} 2\heb{eab}\, 2\heb{ecd} = \g[2]{5} \int_0^{\pi R} \ud y\, \left(\fvb{a} \,\fvb[\prime]{b} - \fvb[\prime]{a} \,\fvb{b} \right) \left(\fvb{c} \,\fvb[\prime]{d} - \fvb[\prime]{c} \,\fvb{d} \right) \,\text{.}
\end{equation}

The~second type of sum contains the~KK index not only in the~couplings, but also in the~mass of the~exchanged vector mode, explicitly
\begin{equation}\eql{E2-sum2}
  \sum_{k \geq 0} m_{\pmb e}^2 \, \gb{eab}\gb{ecd} = \g[2]{5} \int_0^{\pi R}\ud y\, \big(\fvb{a} \,\fvb{b}\big)^\prime \big(\fvb{c} \,\fvb{d}\big)^\prime \,\text{.}
\end{equation}

In order to get all terms of the~invariant matrix element in a~similar form, we need one more formula, which follows directly from the~relation between wave functions and masses \eq{eigenfunctions} and the~coupling definition \eq{couplings-4V}:
\begin{equation}\eql{g4-int}
  4\overline m^2 \gb[2]{abcd} = 2\g[2]{5} \int_0^{\pi R}\ud y\, \Big[ \big(\fvb{a} \,\fvb{b}\big)^\prime \big(\fvb{c} \,\fvb{d}\big)^\prime + \fvb{a} \,\fvb{b} \,\fvb[\prime]{c} \,\fvb[\prime]{d} + \fvb[\prime]{a} \,\fvb[\prime]{b} \,\fvb{c} \,\fvb{d} \Big] \,\text{.}
\end{equation}

Now we gather all the~remaining divergent terms from the~contact four-boson interaction \eq{4V} and $s$, $t$, $u$ channel exchange of vector and scalar modes \eq{st-qq}, \eq{st-diag} and \eq{st-scalar}, employ the~derived sum rules and the~relation \eq{g4-int}.
Interestingly enough, the~resulting invariant matrix element for the~process in question then takes on quite a~simple form
\begin{align}
  \M = \frac s 4 \frac{\g[2]{5}}{m_{\pmb a}m_{\pmb b}m_{\pmb c}m_{\pmb d}} & \left(f^{abe}f^{cde} - f^{ace}f^{bde} + f^{ade}f^{bce}\right) \times {}\notag\\
  \times \Bigg\{ &(1-3\cos\theta) \int_0^{\pi R}\ud y\, \big(\fvb[\prime]{a} \,\fvb{b} \,\fvb{c} \,\fvb[\prime]{d} + \fvb{a} \,\fvb[\prime]{b} \,\fvb[\prime]{c} \,\fvb{d} \big)\notag\\
  {}-{} &(1+3\cos\theta) \int_0^{\pi R}\ud y\, \big(\fvb[\prime]{a} \,\fvb{b} \,\fvb[\prime]{c} \,\fvb{d} + \fvb{a} \,\fvb[\prime]{b} \,\fvb{c} \,\fvb[\prime]{d} \big)\notag\\
  {}-{} &2\cos\theta \int_0^{\pi R}\ud y\, \big(\fvb[\prime]{a} \,\fvb[\prime]{b} \,\fvb{c} \,\fvb{d} + \fvb{a} \,\fvb{b} \,\fvb[\prime]{c} \,\fvb[\prime]{d} \big) \Bigg\} + \Or \,\text{.}
\end{align}

The~whole divergent part of the~matrix element is proportional to the~expression $f^{abe}f^{cde} - f^{ace}f^{bde} + f^{ade}f^{bce}$.
However, this is zero due to the~familiar Jacobi identity.

Thus we conclude that $2 \rightarrow 2$ scattering amplitude of longitudinal gauge bosons contains no terms growing indefinitely with the~energy.
We have shown this fact without any assumptions regarding the~colors or the~KK mode numbers of the~gauge bosons in the~initial and final state.

Note that the~elastic scattering of two identical longitudinal vector modes studied in Ref.~\citen{csaki-interval} is a~special case contained in our general formulae.
Since all the~gauge fields satisfy the~same boundary conditions, the~masses and wave functions (thus, couplings as well) are color-insensitive and are uniquely identified by their KK indices.
This implies that there is no contribution from \eq{st-qq} and \eq{st-scalar} to the~scattering amplitude.
Furthermore, in this special case it is possible to combine \eq{E2-sum2} and \eq{g4-int} to one compact sum rule $\sum_k 3 m_k^2 (\g{nnk})^2 = 4 m_n^2\, \g[2]{nnnn}$.

\section{Conclusions}
We have studied the~gauge sector of a~5D toy model with EWSB triggered by a~non-trivial choice of boundary conditions in the~fifth dimension.
This class of models has already been intensively studied in the~literature, but many authors prefer a~more traditional approach to the~extra dimensions known as {\em orbifolding}~-- one starts with an~infinite extra dimension and compactifies it by a~set of identifications (most commonly to $S^1/Z_2$ orbifold); such a~procedure then implies certain boundary conditions for the~fields.
Another already studied possibility that we have also chosen in this work, is the~interval approach, where one starts straight away with a~finite space interval and then figures out, what the~consistent boundary conditions are.

We have derived the~set of consistent boundary conditions for a~simple model with $SU(2)$ gauge symmetry solely from the~principle of least action and the~requirement of gauge independence of scattering amplitudes.
Any choice belonging to this set leads to the~theory with well-behaved scattering amplitudes of longitudinal vector bosons, i.e. all terms growing as positive power of energy cancel out.
This was explicitly demonstrated\footnote{The~technical details of all the~calculations may be found in Ref.~\citen{diplo}.} on a~general $2 \rightarrow 2$ scattering process without any assumptions regarding the~colors or KK mode numbers of the~gauge bosons in the~initial and final state (and without relying on the KK equivalence theorem).
Previously published results of other authors (see Refs.~\citen{csaki-tasi,csaki-interval,kk-unitarity,deconstruction-sum-rules,kk-gf,kk-transformations}) covered only certain special cases of this model (e.g. a~special choice of boundary conditions, or the~discussion of an~elastic scattering process only).
Our present work is therefore an~improvement and generalization of these earlier results.

\section*{Acknowledgments}
The work was supported by the grant of the Ministry of Education of the Czech Republic MSM 0021620859.

\end{document}